\documentclass[amsmath,amssymb,superscriptaddress,nobalancelastpage,prx,twocolumn]{revtex4-1}

\usepackage{graphicx}
\usepackage{varioref}
\usepackage{xr-hyper}
\usepackage{xcolor}
\usepackage{nicefrac}
\usepackage{xfrac}
\usepackage{hyperref}
\hypersetup{colorlinks,linkcolor=blue,urlcolor=blue,citecolor=blue}
\usepackage[normalem]{ulem}
\usepackage{lineno}
\usepackage{amsmath} 
\usepackage{amssymb}
\usepackage{titlesec}

\begin{document}

\title{Quantum Critical Magnetic Excitations in Spin-1/2 and Spin-1 Chain Systems}  

\author{Y. Xu}
\email{yxu@phy.ecnu.edu.cn}
\affiliation{State Key Laboratory of Surface Physics, and Department of Physics, Fudan University, Shanghai 200438, China}
\affiliation{Key Laboratory of Polar Materials and Devices (MOE), School of Physics and Electronic Science, East China Normal University, Shanghai 200241, China}

\author{L. S. Wang}
\affiliation{State Key Laboratory of Surface Physics, and Department of Physics, Fudan University, Shanghai 200438, China}

\author{Y. Y. Huang}
\affiliation{State Key Laboratory of Surface Physics, and Department of Physics, Fudan University, Shanghai 200438, China}

\author{J. M. Ni}
\affiliation{State Key Laboratory of Surface Physics, and Department of Physics, Fudan University, Shanghai 200438, China}

\author{C. C. Zhao}
\affiliation{State Key Laboratory of Surface Physics, and Department of Physics, Fudan University, Shanghai 200438, China}

\author{Y. F. Dai}
\affiliation{State Key Laboratory of Surface Physics, and Department of Physics, Fudan University, Shanghai 200438, China}

\author{B. Y. Pan}
\affiliation{State Key Laboratory of Surface Physics, and Department of Physics, Fudan University, Shanghai 200438, China}
\affiliation{School of Physics and Optoelectronic Engineering, Ludong University, Yantai, Shandong 264025, China}

\author{X. C. Hong}
\affiliation{State Key Laboratory of Surface Physics, and Department of Physics, Fudan University, Shanghai 200438, China}

\author{P. Chauhan}
\affiliation{The Institute for Quantum Matter, Department of Physics and Astronomy, The Johns Hopkins University, Baltimore, Maryland 21218, USA}

\author{S. M. Koohpayeh}
\affiliation{The Institute for Quantum Matter, Department of Physics and Astronomy, The Johns Hopkins University, Baltimore, Maryland 21218, USA}
\affiliation{Department of Materials Science and Engineering, The Johns Hopkins University, Baltimore, Maryland 21218, USA}

\author{N. P. Armitage}
\affiliation{The Institute for Quantum Matter, Department of Physics and Astronomy, The Johns Hopkins University, Baltimore, Maryland 21218, USA}

\author{S. Y. Li}
\email{shiyan\_li@fudan.edu.cn}
\affiliation{State Key Laboratory of Surface Physics, and Department of Physics, Fudan University, Shanghai 200438, China}
\affiliation{Collaborative Innovation Center of Advanced Microstructures, Nanjing 210093, China}
\affiliation{Shanghai Research Center for Quantum Sciences, Shanghai, 201315, China}

\date{\today}

\begin{abstract}
The study of CoNb$_2$O$_6$ sits at the confluence of simplicity and complexity: on one hand, the model for Ising chains---the building blocks of CoNb$_2$O$_6$---in a transverse field, can be exactly solved and, thus, serves as an archetype of quantum criticality; on the other hand, the weak but nonzero interchain coupling adds geometric frustration to the stage, substantially complicating the phase diagram. Here we utilize low-temperature specific heat and thermal conductivity measurements to study the low-lying magnetic excitations in CoNb$_2$O$_6$ and its spin-1 analogue NiNb$_2$O$_6$. The thermal conductivity is found to be suppressed around the quantum critical point, where the specific heat is enhanced due to gapless magnetic excitations, pointing to the localized nature of the latter. These results highlight the predominant role of frustration in determining the quantum critical magnetic excitations of spin chains, which may furthermore underlie the remarkable similarities between the phenomenology of these spin-1/2 and spin-1 systems.    
\end{abstract}

\maketitle

\section{INTRODUCTION}

Quantum aspects are particularly evident in low-dimensional systems, leading to substantial corrections to classical pictures. In the context of magnetism, excitations with fractionalized quantum number (e.g., spinons)~\cite{Senthil_Z2_2000,Senthil_Fractionalization_2001} and the Haldane phase~\cite{Haldane_gap,senthil_symmetry-protected_2015} are among the most well-known examples. Under specific circumstances, the low dimensionality can lead to simplification. For instance, by mapping to a spinless fermion system, the transverse-field Ising chain (TFIC) model can be exactly solved~\cite{lieb_two_1961,pfeuty_one-dimensional_1970}. The closest realization of this model has long been thought to be CoNb$_2$O$_6$ under a transverse field~\cite{zamolodchikov_integrals_1989,PhysRevB.60.3331,lee_interplay_2010,Co_neutron,Co_NMR,Cabrera_excitations_2014,Robinson_quasiparticle_2014,Co_THz,liang_heat_2015,Co-THZ-PRB,morris_duality_2021}. 

The physics of CoNb$_2$O$_6$, however, extends far beyond: when isolated, the spin-1/2 Ising chains, as the building blocks, order ferromagnetically only at zero temperature, and its evolution under a transverse field can be described elegantly by the TFIC model, featuring a transition to paramagnetic state at an one-dimensional (1D) quantum critical point (QCP)~\cite{sachdev_quantum_1999}; on the flip side, CoNb$_2$O$_6$ being quasi-1D means that these Ising chains are weakly coupled to give rise to 3D magnetic phases at finite temperatures, which terminate at a 3D QCP slightly above the 1D one~\cite{scharf_magnetic_1979,hanawa_anisotropic_1994,Co_NMR}. The arrangement of the Ising chains on a triangular lattice brings in geometric frustration, which is notoriously challenging to tackle~\cite{wannier_antiferromagnetism_1950}. Such a Janus-faced nature of CoNb$_2$O$_6$ has led to several beautiful demonstrations: while subtle properties of 1D Ising quantum critical physics were unveiled by spectroscopy~\cite{zamolodchikov_integrals_1989,Co_neutron,Co_THz,Cabrera_excitations_2014,Robinson_quasiparticle_2014,Wan_Resolving_2019,Co-THZ-PRB,morris_duality_2021}, a plethora of magnetic phases were theoretically predicted as a result of the interplay between quantum criticality and frustration~\cite{lee_interplay_2010}. Moreover, it was shown recently that the frustration-induced quantum motion of domain walls can be described by a model of twisted Kitaev chain, reminiscent of the honeycomb Kitaev spin liquid~\cite{morris_duality_2021}.

Central to these studies is the detection of the magnetic excitations across the QCP~\cite{Co_NMR,Co_neutron,Co_THz,Cabrera_excitations_2014,Robinson_quasiparticle_2014,liang_heat_2015,Co-THZ-PRB,Wu_arxiv}. In low-energy probes like nuclear magnetic resonance, the quasielastic mode associated with domain-wall quasiparticles (kinks) was found to disappear above the 1D QCP~\cite{Co_NMR}. Neutrons create pairs of kinks in the ordered phase, which were observed to change character to spin-flips in the paramagnetic phase~\cite{Co_neutron}. Most prominently, around the 1D QCP, discrete modes were observed both by neutron and time-domain terahertz (THz) spectroscopy, with the energy ratio of the two lowest-lying ones approaching the golden mean predicted for the E8 spectrum, with E8 referring to the maximum exceptional Lie algebra~\cite{zamolodchikov_integrals_1989,Co_neutron,Co_THz,Co-THZ-PRB}. In this context, the bound pairs of kinks can be viewed as the solid-state version of mesons composed of a quark and an antiquark~\cite{zamolodchikov_integrals_1989}. Specific heat study revealed a significant band of gapless fermionic magnetic excitations around the 3D QCP, constituting a large portion of the total spin density of states~\cite{liang_heat_2015}. These gapless excitations were argued to be distinct from the discrete modes evident in spectroscopy~\cite{liang_heat_2015}. The comprehensive understanding of the magnetic excitations around the QCPs in CoNb$_2$O$_6$ out of the results from various probes remains to be understood.

Low-temperature thermal conductivity measurement has proven to be a powerful means in the study of magnetic excitations in low-dimensional magnets~\cite{Sologubenko_thermal,Sologubenko_heat,sologubenko_universal_2003,Hess-magnonPRB,Hess-magnonPRL,Kordonis_spin,Sologubenko_Magnetothermal_2007,Sologubenko_Field_2008,Sologubenko_evidence_2009,yamashita_thermal-transport_2009,yamashita_highly_2010,YMGO,Yu_heat_2017,Leahy_anomalous_2017,Yu_ultra_2018,Ni_absence_2019,Hope_thermal_2019,Hentrich_unusual_2018,li_possible_2020,Pan_specific_2021,rao_survival_2021,huang_heat_2021}. Here, we utilize such measurements down to 60 mK to investigate the quantum critical magnetic excitations in CoNb$_2$O$_6$. We observed an absence of direct contribution to the thermal conductivity from the gapless fermionic magnetic excitations evident in specific heat around the QCP. Furthermore, the thermal conductivity was found to be diminished in the field range where specific heat is enhanced. These results are interpreted as the scattering of heat-carrying phonons off localized magnetic excitations around the QCP, signalling a prominent role of frustration in determining these quantum critical magnetic excitations. A similar set of phenomenology was also observed in the quasi-1D spin-1 Heisenberg ferromagnetic chain system NiNb$_2$O$_6$. The similarity between the two compounds is remarkable, considering the apparently distinct settings of spin quantum number, patterns of long-range order, and magnetic anisotropy. 

\section{METHODS}

\begin{figure}
\begin{center}
 		\includegraphics[width=0.45\textwidth]{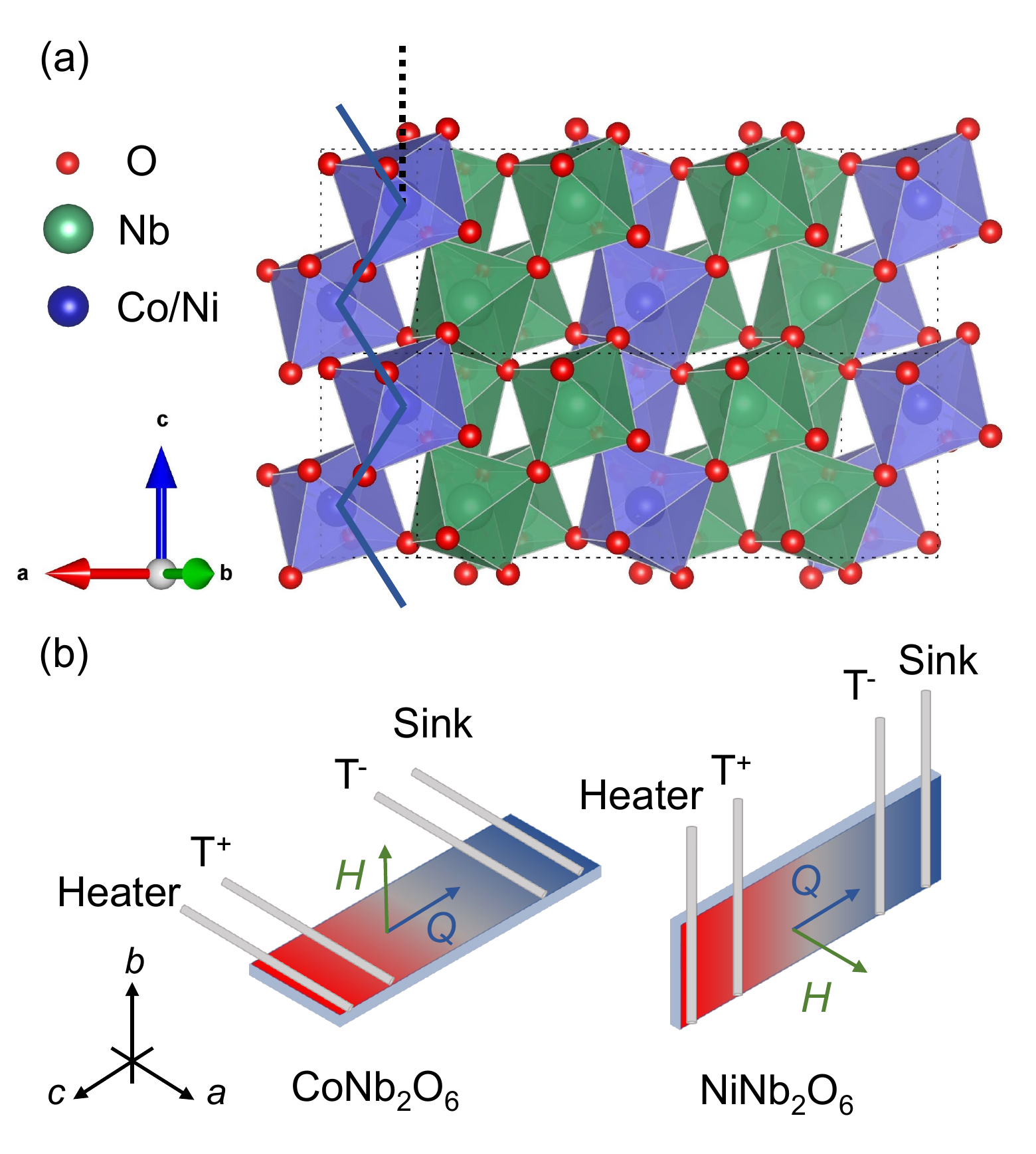}
 	\end{center}
\centering
\caption{ (a) The crystal structure of CoNb$_2$O$_6$ and NiNb$_2$O$_6$. The blue solid line highlights the zigzag chain formed by Co/Ni spins. Note that there is a finite canting angle between the local easy axis (not shown) and the $c$ axis (thick dotted line), and the direction of the former not to be confused with the zigzag chain (thick solid line). (b) The experimental setup for thermal conductivity measurements for CoNb$_2$O$_6$ and NiNb$_2$O$_6$. The magnetic field was applied along $b$ for CoNb$_2$O$_6$ and along $a$ for NiNb$_2$O$_6$.} 
\end{figure}

\begin{figure*}
\begin{center}
 		\includegraphics[width=1\textwidth]{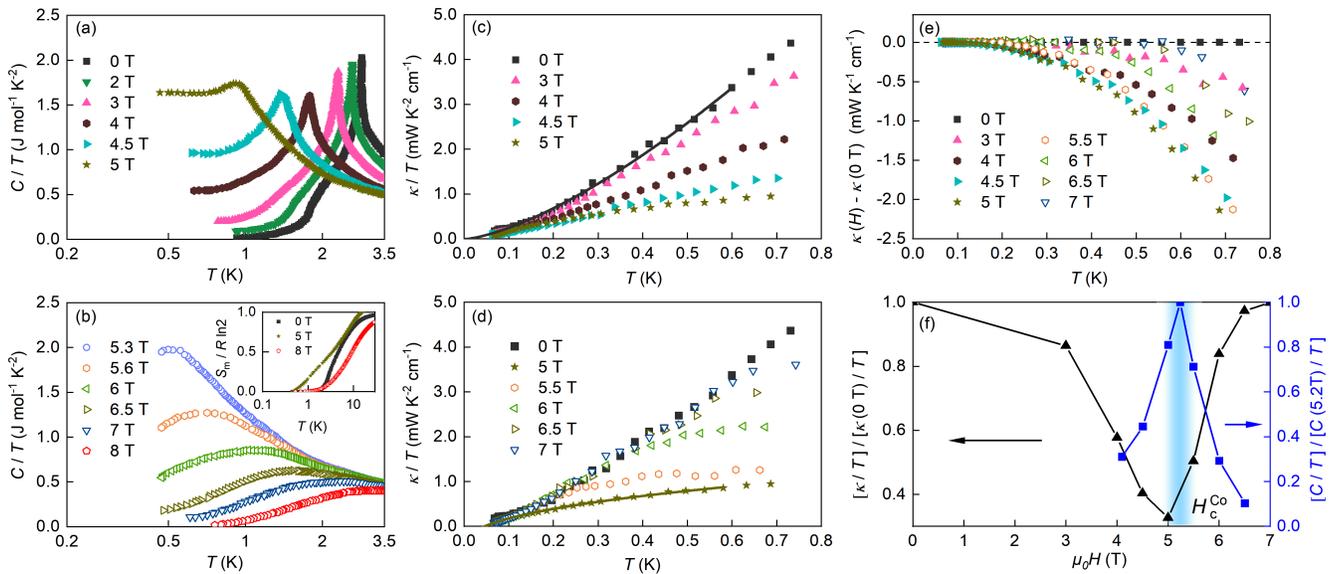}
 	\end{center}
\centering
\caption{(a,b) The specific heat $C/T$ of CoNb$_2$O$_6$ below and above $\mu_0 H^{\rm Co}_{\rm c} \sim$ 5.24 T, respectively. The inset of (b) shows the magnetic entropy $S_{\rm m}$ of CoNb$_2$O$_6$ at selected fields. The data in these two panels were adapted from Ref.~\cite{liang_heat_2015}. (c,d) The thermal conductivity $\kappa/T$ of CoNb$_2$O$_6$ at fields below and above $H^{\rm Co}_{\rm c}$, respectively. The solid lines are the fitting of the 0 and 5 T data to $\kappa/T$ = $a$ + $bT^{\alpha -1}$. (e) The influence of magnetic excitations on the thermal conductivity reflected in $\Delta \kappa$ $\equiv$ $\kappa (H)~-$ $\kappa (0~\rm T)$. (f) The magnetic field evolution of the normalized thermal conductivity and magnetic specific heat at 0.45 K. The cyan shaded region depicts the position of $H^{\rm Co}_{\rm c}$.} 
\end{figure*}

As illustrated in Fig. 1(a), both columbite CoNb$_2$O$_6$ and NiNb$_2$O$_6$ crystallize in the space group $Pbcn$, consisting of zigzag chains of edge-sharing octahedra along the crystallographic $c$ axis with ferromagnetic interactions between nearest-neighbor effective spin-1/2 Co$^{2+}$ or spin-1 Ni$^{2+}$ ions, respectively, and the interchain coupling is considerably smaller~\cite{heid_magnetic_1995,Ni_Magnetization}. In CoNb$_2$O$_6$, a strong single-ion anisotropy due to crystal field effects from the distorted CoO$_6$ octahedra forces the Co$^{2+}$ spins to be Ising aligned along the local easy axis, which lies canted from the $ac$ plane in an alternating fashion from bond to bond and slightly cants away from the $c$ axis~\cite{heid_magnetic_1995,cant1,cant2,cant3,morris_duality_2021}. In the $ab$ plane, the weakly coupled Ising chains sit on a triangular lattice, a prototypical motif for geometric frustration~\cite{wannier_antiferromagnetism_1950}. The magnetism in NiNb$_2$O$_6$ is more isotropic, so that the exchange coupling can be described by a Heisenberg term, with an additional term for the uniaxial anisotropy~\cite{heid_magnetism_1996,THz}. Frustration has also been argued to manifest itself in NiNb$_2$O$_6$~\cite{heid_magnetism_1996}.

Single crystals of CoNb$_2$O$_6$ and NiNb$_2$O$_6$ were grown at Fudan University and the Johns Hopkins University, respectively, by the floating zone method and oriented by Laue diffraction. The specific heat of NiNb$_2$O$_6$ was measured by the relaxation method in a Quantum Design physical property measurement system equipped with a dilution refrigerator. For the thermal conductivity measurements, the CoNb$_2$O$_6$ and NiNb$_2$O$_6$ samples were cut into a rectangular shape of dimensions 2.0 $\times$ 0.5 $\times$ 0.5 mm$^3$ and 2.1 $\times$ 0.3 $\times$ 0.1 mm$^3$, respectively, with the longest dimension along the $c$ axis. Four silver wires were attached to the sample with silver paint for thermal conductivity measurements with a heat current along $c$. The thermal conductivity was measured in a dilution refrigerator, using a standard four-wire steady-state method with two RuO$_2$ chip thermometers, calibrated $in$ $situ$ against a reference RuO$_2$ thermometer. Figure 1(b) gives a schematic of the experimental setup for thermal conductivity measurements, with a heat current along the chain direction $c$ and transverse magnetic fields along $b$ for CoNb$_2$O$_6$ and along $a$ for NiNb$_2$O$_6$, respectively. The specific heat of NiNb$_2$O$_6$ was also measured with the magnetic field along the $a$ axis.

\section{RESULTS AND DISCUSSIONS}

\subsection{Magnetic excitations around the QCP in CoNb$_2$O$_6$}

The specific heat $C$ data extracted from Ref.~\cite{liang_heat_2015} and our thermal conductivity $\kappa$ data for CoNb$_2$O$_6$ are compiled in Fig. 2. As shown in Fig. 2(a), the 3D magnetic order onsets at 2.85 K at 0~T, leading to a sharp peak in the temperature dependence of $C/T$, which is gradually suppressed by a transverse magnetic field~\cite{liang_heat_2015}. By tracking the peak position in the field dependence of $C/T$, the critical field where the 3D magnetic order is fully suppressed was estimated to be $\mu_0 H^{\rm Co}_{\rm c} \sim$ 5.24 T ($\mu_0$ being the vacuum permeability)~\cite{liang_heat_2015}. The most prominent feature of the specific heat data is the low-temperature plateau in $C/T$ for fields slightly below $H^{\rm Co}_{\rm c}$, as exemplified by the 5~T data in Fig. 2(a). Such a plateau in $C/T$, i.e., a linear term in $C$ is characteristic of gapless fermionic excitations. In an insulator like CoNb$_2$O$_6$, such excitations are presumably from the magnetic degrees of freedom. These excitations constitute a large portion ($\sim$ 30\%) of the total spin density of states~\cite{liang_heat_2015}, as evident by the larger value of both the low-temperature $C/T$ [Figs. 2(a) and 2(b)] and magnetic entropy $S_{\rm m}$ [inset of Fig. 2(b)] at fields around $H^{\rm Co}_{\rm c}$ than at lower and higher fields.

The thermal conductivity of an insulating magnet can generally be decomposed as $\kappa$ = $\kappa_{\rm m} + \kappa_{\rm p}$, in which the two terms represent the contribution from magnetic excitations and phonons, respectively. If the magnetic excitations are gapless and fermionic, then in the simplest case (free fermions), they are expected to give rise to a linear contribution to $\kappa$, i.e., $\kappa_{\rm m} \sim T$~\cite{ziman_electrons_nodate}. At low temperatures ($T <$ 1 K), phonons are usually scattered solely by the sample boundaries, so that $\kappa_{\rm p} \sim$ $T^{\alpha}$, where $\alpha$ is usually between 2 and 3~\cite{PhysRevB.67.174520,magnon_kappa,ziman_electrons_nodate}. Combining these two terms, one can fit the data to 
\begin{equation}
\kappa/T = a + bT^{\alpha -1},    
\end{equation}
and obtain a nonzero residual linear term $\kappa_0/T \equiv \kappa_{\rm m}/T~(T \rightarrow 0) = a$ from the gapless fermionic magnetic excitations, provided they are itinerant. Note that in some exotic cases, the magnetic excitations are not well-defined Landau quasiparticles---as exemplified by spinons in the presence of gauge fluctuations, and their thermal conductivity is expected to be sublinear and, again, one would get a finite $\kappa_0/T$ when Eq. 1 is adopted~\cite{PhysRevLett.95.036403,PhysRevB.76.235124}. Therefore, one would expect (i) $\kappa$ to be enhanced in the same field range where $C$ is enhanced, and (ii) finite values of $\kappa_0/T$ in this field range.

However, as shown in Figs. 2(c) and 2(d), neither feature was observed in our thermal conductivity data. In fact, $\kappa$ shows an exactly opposite evolution with field as compared to $C$: $\kappa$ is seen to be reduced at field values where $C$ is enhanced. This is better visualized in Figs. 2(e) and 2(f), where $\Delta \kappa$ $\equiv$ $\kappa (H)~-$ $\kappa (0~\rm T)$, and the field evolution of the normalized value of specific heat and thermal conductivity, are plotted, respectively. The fitting to Eq. 1 yields $a$ = $-$ 0.01 $\pm$ 0.02 mW K$^{-2}$ cm$^{-1}$ and $\alpha$ = 2.45 $\pm$ 0.04 for 0~T, and $a$ = $-$ 0.44 $\pm$ 0.05 mW K$^{-2}$ cm$^{-1}$ and $\alpha$ = 1.43 $\pm$ 0.03 for 5~T. At 0~T, the only candidate for heat carriers is the phonons. Indeed, $\kappa_0/T$ is essentially negligible, and the value of $\alpha$ is consistent with the boundary scattering mechanism for phonons. At 5~T where the contribution to $C$ from gapless magnetic excitations is most pronounced, the fitting of $\kappa$ gives, surprisingly, an unphysical negative $\kappa_0/T$ and an $\alpha$ considerably smaller than the 0 T value and abnormally lower than 2. 

\begin{figure*}
\begin{center}
 		\includegraphics[width=1\textwidth]{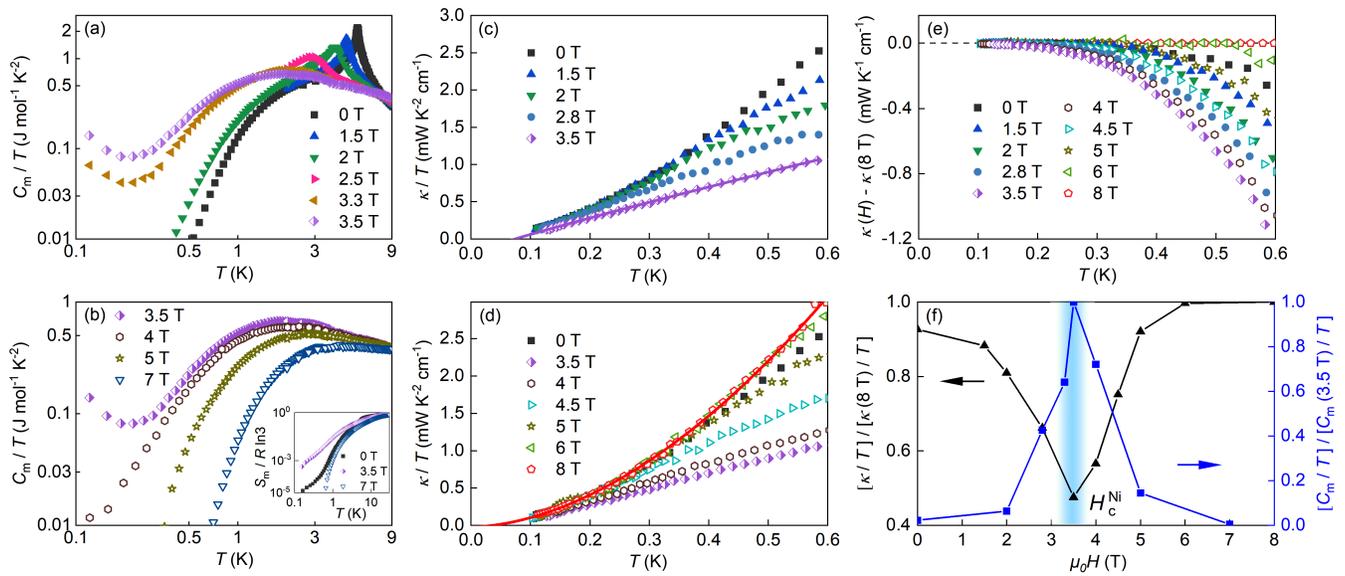}
 	\end{center}
\centering
\caption{(a,b) The magnetic specific heat $C_{\rm m}/T$ of NiNb$_2$O$_6$ below and above $\mu_0 H^{\rm Ni}_{\rm c} \sim$ 3.5 T, respectively. The inset of (b) shows the magnetic entropy $S_{\rm m}$ of NiNb$_2$O$_6$ at selected fields. (c,d) The thermal conductivity $\kappa/T$ of NiNb$_2$O$_6$ at fields below and above $H^{\rm Ni}_{\rm c}$, respectively. The solid lines are the fitting of the 3.5 and 8 T data to $\kappa/T$ = $a$ + $bT^{\alpha -1}$. (e) The influence of magnetic excitations on the thermal conductivity reflected in $\Delta \kappa$ $\equiv$ $\kappa (H)~-$ $\kappa (8~\rm T)$. (f) The magnetic field evolution of the normalized thermal conductivity and magnetic specific heat at 0.4 K. The cyan shaded region depicts the position of $H^{\rm Ni}_{\rm c}$.} 
\end{figure*}

The absence of a positive $\kappa_0/T$ clearly rules out the direct contribution to $\kappa$ from the gapless fermionic magnetic excitations around the QCP (see Sec. I in the Supplemental Material~\cite{SM_note} for more discussions). This apparent contradiction with the specific heat data can only be removed if these excitations are localized. Note that, the excitations are deemed as localized either when they are strongly scattered so that they cannot propagate along a certain direction at a considerable length scale (e.g., significantly larger than the interspin distance), or they intrinsically exist in closed loops. However, these excitations do have a substantial impact on $\kappa$: the value of $\kappa$ and $\alpha$ should not change with magnetic field if the heat-carrying phonons are scattered solely by the sample boundary. The considerably smaller value of $\kappa$ and the unusual value of $\alpha$ around $H^{\rm Co}_{\rm c}$ signal the presence of other scattering mechanisms for phonons, with a scattering strength peaking in this field range. These features strongly suggest the additional scattering of phonons by the gapless magnetic excitations. As demonstrated in detail in Ref.~\cite{huang_heat_2021}, both the unphysical negative $\kappa_0/T$ and the unusual values of $\alpha$ are consequences of forcing the fitting with Eq. 1 in systems with heat-carrying phonons scattered by magnetic degrees of freedom. At 5~T, which is slightly below the critical field, $\Delta \kappa$ is finite in almost the entire temperature range [Fig. 2(e)], consistent with the magnetic excitations that scatter phonons being gapless. 

A natural speculation is that the localization of the magnetic excitations is induced by disorder. Localization effects in 1D systems have been investigated mainly in the charge sector~\cite{Giamarchi_quantum_2004}, with only a few studies on the magnetic thermal conductivity for several spin ladder and spin chain systems~\cite{Sologubenko_Magnetothermal_2007,Sologubenko_Field_2008,Sologubenko_evidence_2009,Karahalios_finite_2009}. In the spin ladder compound (C$_5$H$_{12}$N)$_2$CuBr$_4$, the absence of a measurable magnetic thermal conductivity was argued to result from spinon localization within finite ladder segments~\cite{Sologubenko_evidence_2009}. The thermal conductivity is thus dominated by the phonon contribution, which is scattered by the localized spinons via umklapp scattering~\cite{Sologubenko_evidence_2009}. This may seem to be a viable scenario for CoNb$_2$O$_6$. However, this scenario is favored by the extremely weak interladder coupling of (C$_5$H$_{12}$N)$_2$CuBr$_4$ ($\sim$ 27 mK). When the degree of one dimensionality is lowered, as in, e.g., the spin chain system Cu(C$_4$H$_4$N$_2$)(NO$_3$)$_2$, a similar scattering of phonons by localized magnetic excitations was proposed~\cite{Sologubenko_Magnetothermal_2007}. However, instead of tunneling through the strong impurities that cut the chains into segments, the excitations can now bypass the impurities by hopping to the neighbouring chains~\cite{Sologubenko_Magnetothermal_2007}. This leads to a considerable mean free path of the excitations and consequently a finite magnetic thermal conductivity~\cite{Sologubenko_Magnetothermal_2007}: the mean free path reaches $\sim$ 750 $\mu$m in the spin chain system Ni(C$_2$H$_8$N$_2$)$_2$NO$_2$(ClO$_4$), rendering the quasiparticle motion effectively 3D~\cite{Sologubenko_Field_2008}. If at play, such hopping-assisted propagation of the magnetic excitations is expected to give rise to a measurable magnetic thermal conductivity in CoNb$_2$O$_6$, a system much more 3D than the above-mentioned organic compounds, as reflected by its orders of magnitude larger interchain coupling and transition temperatures of the 3D magnetic ordering~\cite{heid_magnetic_1995,PhysRevB.60.3331,lee_interplay_2010,Regnault_inelastic_1994,Lancaster_magnetic_2006,Klanjsek_controlling_2008}. This is in sharp contrast to our observation (see Sec. I in the Supplemental Material~\cite{SM_note} for an estimation of the mean free path).

Therefore, the localization mechanism of the magnetic excitations in CoNb$_2$O$_6$ must be distinct from the typical disorder-induced scenario in 1D spin systems, implying the essential role of the 3D nature and the consequent frustration. In fact, a similar picture of heat carrying phonons scattered by localized magnetic degrees of freedom, has also been discussed for several famous quantum spin liquid candidates like YbMgGaO$_4$~\cite{YMGO}, $\alpha$-RuCl$_3$~\cite{Yu_ultra_2018,Hentrich_unusual_2018}, EtMe$_3$Sb[Pd(dmit)$_2$]$_2$~\cite{Ni_absence_2019,Hope_thermal_2019}, and ZnCu$_3$(OH)$_6$Cl$_2$~\cite{huang_heat_2021}, as a consequence of the combination of disorder and frustration. From an experimental perspective, our finding of the localized nature of the critical excitations, together with the Kitaev physics unveiled by THz spectroscopy, and the glassy response found in a.c. calorimetry away from the QCP~\cite{liang_heat_2015}, highlights the complexity introduced by frustration to the physics of CoNb$_2$O$_6$.

The origin of the fermionic excitations in the ordered phase as found by specific heat measurements is unclear~\cite{liang_heat_2015}. One possibility is the quasiparticles arising from breaking the bound pairs of kinks~\cite{Co_neutron}. In Ref.~\cite{liang_heat_2015}, however, it was argued that the magnetic excitations evident in specific heat should be entirely distinct from those observed by spectroscopy, considering the higher energy scale of the latter: the lowest discrete mode was observed around 1.2~meV at 0~T and 0.4~meV at 5~T, while the plateau in specific heat was observed below 1~K. Such an energy scale argument alone, however, does not necessarily exclude the possibility that the excitations seen in different probes share a similar origin. For an isolated Ising chain, the spectrum for critical kinks is gapless; however, in a quasi-1D magnet like CoNb$_2$O$_6$, the nonzero interchain coupling can be approximated by an effective longitudinal field, which opens a gap and stabilizes the bound states, as observed in spectroscopy~\cite{zamolodchikov_integrals_1989,carr_spectrum,Co_neutron,Co_THz,Co-THZ-PRB}. In this case, the gap is momentum dependent, leaving the gapless feature intact only at specific wave vectors~\cite{lee_interplay_2010}. In THz spectroscopy, the momentum space relevant to the optical responses is inherently limited to the Brillouin zone center. As explicitly stated in Ref.~\cite{Co_neutron}, the neutron scattering measurements presented therein were performed in a scattering plane with an incomplete gap softening: the spectrum was obtained at a wave vector of (3.6(1),0,0), whereas a complete gap softening is only expected at the location of the 3D magnetic long-range order Bragg peaks. Consequently, the critically soft magnetic spectrum may evade the detection of spectroscopic probes and manifest only in specific heat and thermal conductivity measurements, which treat all portions of the Brillouin zone on an equal footing. 

A remaining issue is the field range associated with the critical excitations. Strictly speaking, closest agreement with the E8 mass ratio for the critical excitations is expected at the 1D QCP~\cite{carr_spectrum}. Although the gapless fermionic magnetic excitations were argued to be observed around the 3D QCP in Ref.~\cite{liang_heat_2015}, the distinction is not clear considering the closeness between the 1D and the 3D QCP in CoNb$_2$O$_6$~\cite{Co_neutron,Co_NMR,liang_heat_2015}. In this regard, it would be interesting to study the E8 excitations by specific heat and thermal conductivity measurements in BaCo$_2$V$_2$O$_8$, in which the 1D and 3D QCPs are well separated~\cite{zhang_observation_2020,zou_E8_2021}.

The above discussions about the energy and field scales lead to the inference that it is likely that in CoNb$_2$O$_6$, the magnetic excitations probed by various experimental probes are closely related, and the localized nature of these excitations revealed by our thermal conductivity data may have profound implications on all these previous studies.

\subsection{Magnetic excitations around the QCP in NiNb$_2$O$_6$}

The magnetic specific heat $C_{\rm m}$ of NiNb$_2$O$_6$ under various magnetic fields, obtained by subtracting the phonon specific heat estimated from the nonmagnetic counterpart ZnNb$_2$O$_6$~\cite{hanawa_disappearance_1992}, is shown in Figs. 3(a) and 3(b). At 0 T, a clear peak marks the N\'eel temperature $T_{\rm N}$ $\sim$ 5.5 K. With increasing field, the peak height is reduced and the peak is shifted to lower temperatures. We estimate the critical field where $T_{\rm N}$ is suppressed to zero as $\mu_0 H_{\rm c}^{\rm Ni}$ $\sim$ 3.5 T. Similar to CoNb$_2$O$_6$, around $H_{\rm c}^{\rm Ni}$, $C_{\rm m}$ is significantly larger than that at lower and higher fields. Consequently, $S_{\rm m}$ [inset of Fig. 3(b)] at $H_{\rm c}^{\rm Ni}$ is also seen to be larger than that at any other fields up to 5 K. 

Moreover, at 3.3 and 3.5~T, $C_{\rm m}/T$ exhibits an upturn at the lowest temperatures, reminiscent of a Schottky contribution from nuclear spins. However, nuclear Schottky contribution is expected to increase monotonically with field, inconsistent with our data. The upturn in $C_{\rm m}/T$ may eventually level off and form a plateau below the lowest temperature measured, similar to CoNb$_2$O$_6$. Alternatively, it is also possible that the divergence is intrinsic. In either case, the critical-like divergent behavior of $C_{\rm m}/T$ signals the presence of magnetic degrees of freedom whose gap vanishes around the QCP.

Figures 3(c) and 3(d) show $\kappa/T$ of NiNb$_2$O$_6$ at various fields along the $a$ axis. The results with fields along $b$ are similar, as shown in Sec. II in the Supplemental Material~\cite{SM_note}. Similar to CoNb$_2$O$_6$, $\kappa/T$ is suppressed for field values around $H_{\rm c}^{\rm Ni}$, which is better illustrated in Fig. 3(e), where $\Delta \kappa$ $\equiv$ $\kappa (H)~-$ $\kappa (8~\rm T)$ is plotted, and Fig. 3(f), where the field evolution of $\kappa$ and $C_{\rm m}$ is displayed. Again, as a function of magnetic field, $\kappa$ shows an anticorrelation with $C_{\rm m}$, $i.e.$, $\kappa$ is suppressed whenever $C_{\rm m}$ is enhanced. This immediately indicates that the magnetic degrees of freedom underlying $C_{\rm m}$ do not give a direct net contribution to $\kappa$, but serve as scattering centers for heat-carrying phonons instead. Indeed, the fitting to Eq. 1 yields $a$ = $-$ 0.01 $\pm$ 0.01 mW K$^{-2}$ cm$^{-1}$ and $\alpha$ = 2.80 $\pm$ 0.02 for 8~T, and $a$ = $-$ 0.19 $\pm$ 0.02 mW K$^{-2}$ cm$^{-1}$ and $\alpha$ = 1.91 $\pm$ 0.02 for 3.5~T. $\kappa (3.5~\rm T)$ is suppressed relative to $\kappa (8~\rm T)$ in the whole temperature range of our measurement, consistent with the gapless behavior seen in $C_{\rm m} (3.5~{\rm T})$. Note also the lower value of $\kappa (0~\rm T)$ than $\kappa (6~\rm T)$ and $\kappa (8~\rm T)$ (the latter two almost overlap), indicating the incomplete recovery of the phonon thermal conductivity for the former, which justifies our choice of $\kappa (8~\rm T)$ as the reference data in the definition of $\Delta \kappa$.

To our knowledge, a similar divergent specific heat that only becomes evident at $T<$ $\sim$ 0.3 K with no concomitant thermal conductivity contribution was also observed in, e.g., the spin-triplet superconductor UTe$_2$, where it was attributed to a well-localized density of states or fermionic carriers that are heavily scattered~\cite{UTe2}.

\subsection{Similarities between the phenomenology of CoNb$_2$O$_6$ and NiNb$_2$O$_6$}

The similar set of phenomenology in CoNb$_2$O$_6$ and NiNb$_2$O$_6$, i.e., (i) an anticorrelation between the specific heat and thermal conductivity, with the former enhanced and the latter suppressed around the QCP, and (ii) an absence of the residual linear term and the unusual temperature scaling of the phonon thermal conductivity, point to a unified picture featuring localized critical magnetic excitations.

Although CoNb$_2$O$_6$ and NiNb$_2$O$_6$ share a similar structure, the similarities between the critical excitations are remarkable, considering the vast difference in their spin Hamiltonian. To obtain insights on a qualitative level, we consider here the simplest case of isolated chains. The spin Hamiltonian of CoNb$_2$O$_6$ under a transverse field is then:

\begin{equation}
H=-|J| \sum_{\langle i ; j \rangle} {S}_{i}^z {S}_{j}^z-gH \sum_{i} {S}_{i}^x.    
\end{equation}
The spin Hamiltonian of NiNb$_2$O$_6$ under a transverse field can be described by:
\begin{equation}
H=-|J| \sum_{\langle i ; j \rangle} \vec{S}_{i} \cdot \vec{S}_{j}-|D| \sum_{i}\left(S_{i}^{z}\right)^{2}-gH \sum_{i} {S}_{i}^x.
\end{equation}
Compared to Eq. 2, Eq. 3 features (a) spin-1 operators $S_i$ instead of spin-1/2, (b) a more isotropic exchange term $\vec{S}_{i} \cdot \vec{S}_{j}$, and (c) an additional term for the local onsite anisotropy $D$. While the TFIC model in Eq. 2 can be studied analytically, the Hamiltonian for NiNb$_2$O$_6$ has been studied numerically by exact diagonalization and density matrix renormalization group methods~\cite{THz}. Some `in-between' situations of these two distinct models, including the Blume-Capel model, which deal with (a) and (c) but with an Ising exchange term and without the coupling to external field, can also be rigorously solved~\cite{BC1,BC2,Yang_rigorous,Yang_exact}. In these models, compared to spin-1/2, the major consequence of the spin-1 nature is the existence of $S_z$ = 0 states called holes. The holes separate the system into many independent segments of interacting $S$= 1/2 spins. The interplay between the hole excitations and the fermionic excitations within each spin-1/2 Ising segment hence determines the low temperature behavior and the quantum criticality~\cite{Yang_rigorous,Yang_exact}. 

As discussed above, the physics of CoNb$_2$O$_6$ and NiNb$_2$O$_6$ goes well beyond the description of Eqs. 2 and 3, mainly as a result of the nonzero interchain coupling and the resultant frustration. The localized nature of the critical magnetic excitations in both compounds underscores the overarching role of the concerted effort of disorder and frustration in spin chain systems, and low dimensional magnets in a broader sense. As mentioned, the localized nature of the magnetic excitations has been concluded from transport studies on various quantum magnets~\cite{Sologubenko_Magnetothermal_2007,Sologubenko_evidence_2009,YMGO,Yu_ultra_2018,Hentrich_unusual_2018,Ni_absence_2019,Hope_thermal_2019,huang_heat_2021}. However, due to the presence of disorder, frustration, and quantum many-body effects, a clear correspondence between the localized nature of these excitations and their spectroscopic features is still lacking. How the localized excitations manifest themselves in spectroscopy is a challenging issue to be addressed in the future.\\

\section{SUMMARY}

We measured the thermal conductivity of the Ising chain system CoNb$_2$O$_6$. We found an anticorrelation between the transverse field dependence of the thermal conductivity and that of the specific heat, the latter argued to be enhanced due to the quantum critical magnetic excitations which are gapless and fermionic. Contrary to the expectation that these excitations would contribute to a similar enhancement of the thermal conductivity, the heat current is solely carried by phonons which, however, are strongly scattered by the magnetic excitations. We argue that various experimental probes likely measure quantum critical excitations of a common origin, whose localized nature is unveiled for the first time by our thermal conductivity measurements. A similar set of phenomenology was also found in the spin-1 chain NiNb$_2$O$_6$, which likewise undergoes quantum transitions under a transverse field. The loss of mobility of the quantum critical excitations in two systems with magnetic properties distinct in many aspects, highlights a role of frustration not inferior to quantum criticality itself in spin chain systems.

\section*{ACKNOWLEDGEMENTS}

We thank Y. Chen, Y. Zhou and J. Wu for helpful discussions. This work was supported by the Natural Science Foundation of China (Grant No. 12034004), the Ministry of Science and Technology of China (Grant No. 2016YFA0300503), and the Shanghai Municipal Science and Technology Major Project (Grant No. 2019SHZDZX01). Y. Xu was sponsored by the Shanghai Pujiang Program (Grant No: 21PJ1403100) and the Natural Science Foundation of Shanghai (Grant Nos: 21JC1402300 and 21ZR1420500). Work at the Johns Hopkins University was supported as part of the Institute for Quantum Matter, an EFRC funded by the DOE BES under DE-SC0019331. 

Y. Xu and L. S. Wang contributed equally to this work.

\end{document}